\documentclass[%
 reprint,
 amsmath,amssymb,
 aps,
]{revtex4-2}
\usepackage{float}
\usepackage{graphicx}
\usepackage{dcolumn}
\usepackage{bm}
\usepackage[mathlines]{lineno}

\begin{document}

\preprint{APS/123-QED}

\title{Discovery of the Superconductivity in 3d Element Ni doped IrTe$_2$}
\author{Qiang Jing$^{1,2}$}
\email{jingqiang@sdut.edu.cn}
\author{Ping Li$^{4}$}
\author{Yunze Long$^{5}$}
\author{Jingfeng Wang$^{2}$}
\author{Fei Jiao$^{7}$}
\author{Xiaoxiong Wang$^{5}$}
\author{Xiaomin Cui$^{2,6}$}
\author{Wenxiang Jiang$^{1}$}
\author{Guohua Wang$^{1}$}
\author{Yunlong Li$^{1}$}
\author{Gan Liu$^{8}$}
\author{Cao Wang$^{2}$}
\author{Bo Liu$^{2}$}
\author{Dong Qian$^{1,3}$}%
\email{dqian@sjtu.edu.cn}

\affiliation{$^{1}$Key Laboratory of Artificial Structures and Quantum Control (Ministry of Education), Shenyang National Laboratory for Materials Science, School of Physics and Astronomy, Shanghai Jiao Tong University, Shanghai 200240, China}%
\affiliation{$^{2}$Laboratory of Functional Molecules and Materials, School of Physics and Optoelectronic Engineering, Shandong University of Technology, 266 Xincun Xi Road, Zibo, 255000, China}%
\affiliation{$^{3}$Tsung-Dao Lee Institute, Shanghai Jiao Tong University, Shanghai 200240}%
\affiliation{$^{4}$Anhui Jianzhu University, School of Mathematics and Physics, Hefei 230601, China}%
\affiliation{$^{5}$College of Physics, Qingdao University, No.308 Ningxia Road, Qingdao 266071, China}%
\affiliation{$^{6}$Laboratory of Solid State Microstructures and Innovation Center of Advanced Microstructures,  Nanjing University, Nanjing 210093, China). }%
\affiliation{$^{7}$School of Materials Science and Engineering, Shandong University of Technology, 266 Xincun Xi Road, Zibo, 255000, China}%
\affiliation{$^{8}$National Laboratory of Solid State Microstructures, Department of Physics and Collaborative Innovation Center of Advanced Microstructures, Nanjing University, 22 Hankou Road, Nanjing 210093, People's Republic of China}%

\date{\today}

\begin{abstract}
IrTe$_2$ with large spin-orbital coupling (SOC) shows a CDW-like first order structural phase transition from high-temperature trigonal phase to low-temperature monoclinic phase at 270 K, accompanying with a large jump in transport and magnetic measurement as well as in heat capacity. Here, the 3d element Ni has been doped into IrTe$_2$ by growing  Ir$_{1-x}$Ni$_x$Te$_2$ single crystals. Both XRD and XPS results reveal that the Ni atoms have substituted for Ir, which is consistent with the calculation result. Like the CDW behaviour, the structural phase transition shows competition and coexistence with the superconductivity. The monoclinic phase transition has been suppressed gradually with the increase of  the doping amount of Ni, at last giving rise to the stabilization of the trigonal phase with superconductivity. Within 0.1$\leq$x$\leq$0.2, Ir$_{1-x}$Ni$_x$Te$_2$ shows the superconductive behaviour with T$_c$ around 2.6K. The superconductivity shows anisotropy with dimensionless anisotropy parameter $\gamma$=$\xi_{//}$$/$$\xi_{\perp}$$\sim$ 2. Even Ni element shows ferromagnetic behaviour, Ir$_{1-x}$Ni$_x$Te$_2$ only shows weak paramagnetism, no ferromagnetic order is observed in it, which is coincident with the calculation result that their up and down spin density of states compensate each other well. In addition, for other 3d elements Fe, Co and Mn doped IrTe$_2$, only Ir$_{1-x}$Mn$_x$Te$_2$ owns magnetism with magnetic moment of 3.0$\mu$$_B$ to the supercell, theoretically.

\textit{Keywords}: IrTe$_2$; structural phase transition; superconductivity; magnetic elements doping
\end{abstract}

\maketitle


\section{Introduction}

One of major challenges in condensed matter physics is to comprehend the interaction between superconductivity and other modulated electronic ground states, e.g., magnetism, charge/orbital ordering, structure distortion, spin/charge-density wave (SDW/CDW)\cite{1}. Charge density wave (CDW) is a kind of periodic electronic charge modulation, which is accompanied by a distortion of the underlying lattice with the same periodicity \cite{2,3}. It is a collective state that often appears in low-dimensional electronic systems. In many cases it is also related with superconductivity, another collective phenomenon\cite{4}. The driving force of a CDW could be a Fermi surface nesting instability, the formation of local bound states\cite{5,6} or the indirect Jahn-Teller effect\cite{7}. For the interplay between CDW and superconductivity, there exists competition and coexistence of CDW and superconductivity in quasi-two-dimensional transition-metal dichalcogenides \cite{16,17,18}. When the CDW phase is suppressed, the superconductivity emerges. It is believed that this phenomenon is related with high-T$_c$ superconductivity \cite{19,20,21,22,23}. For the interplay between magnetism and superconductivity, it was believed that elements with a large magnetic moment were harmful to the emergence of superconductivity because the magnetism originating from the static ordering of magnetic moments would compete with superconductivity, which is realized through the formation of Cooper pairs. However, the pairing process is a kind of dynamic process realized by two conducting electrons with opposite spin\cite{34}. Spin-orbital coupling (SOC) is proportional to Z$^4$, where Z is the atomic number. Large SOC can result in unique quantum states such as J$_{eff}$=1/2 Mott insulators\cite{9}, topological insulator\cite{10,11}. Furthermore, nonconventional superconductivity pairing can be present in the so-called topological superconductors with large SOC\cite{12,13,14,15}. For IrTe$_2$, both Ir and Te have large atomic number, which makes it own large SOC.
In another aspect, IrTe$_2$ undergoes a CDW-like first order structural phase transition at about T$_s$= 270 K to a phase with a charge-ordered state characterized by a wave vector q$_n$ = [1/(3n+2), 0, 1/(3n+2)] with respect to the room-temperature unit-cell vectors\cite{36,37}. The phase transition is accompanied by a change of its symmetry from trigonal to monoclinic \cite{37}. A large jump appears in transport and magnetic measurement \cite{37}as well as in heat capacity \cite{38}. Later, it was suggested that the structural phase transition was driven by the energy gain due to Ir and Te dimerization, with Ir orbitals playing a key role \cite{39,40,41}. For the superconductive property of IrTe$_2$, IrTe$_2$ thin flake shows superconductive behaviour with T$_c$=3.2K. It was suggested that the CDW-like normal-to-striped charge order transition was suppressed in the thin flake, resulting in the metastable superconductivity\cite{35}. Through element doping (Pd, Pt, Rh and Cu) \cite{29,30,31,32,33}, IrTe$_2$ can become a superconductor too. Then, it offers an appropriate platform to investigate structure/charge/orbital fluctuations and superconductivity under strong SOC. Here, we try to add another physical quantity i.e. ferromagnetism to IrTe$_2$ system to investigate the interaction between structure/charge/orbital fluctuations, superconductivity, strong SOC and ferromagnetism. J.-Q. Yan et al. has investigated the doping effect of several 3d elements by synthesizing M$_{0.5}$IrTe$_2$ (M = Mn, Fe, Co, and Ni) polycrystals. No superconductivity was observed down to 1.80 K in any of the doped compositions. Furthermore, no phase transition was observed and the trigonal structure was stable down to 1.8 K\cite{51}.  Here, experimentally, the doping effect of nickel on the properties of IrTe$_2$ is investigated.  Differently from M$_x$TiSe$_2$ (M = Cr, Mn, Fe, Ni), whose CDW phase transition disappears at x$\sim$0.1, whereas when x$\geq$0.25, the CDW state reappears \cite{52},  the suppression on the phase transition of IrTe$_2$ strengthens proportionally to the doping amount and no
reappearance of the phase transition is observed within x $\leq$ 0.2. The superconductivity  is observed, while no ferromagnetism is observed even Ni is a kind of ferromagnetic element. Theoretically, the doping effect of Mn, Fe, Co, and Ni on the magnetic property of IrTe$_2$ is investigated too.

\section{Experimental and calculational details}
\begin{figure*}[t]
\centering
\includegraphics[width=17.2cm]{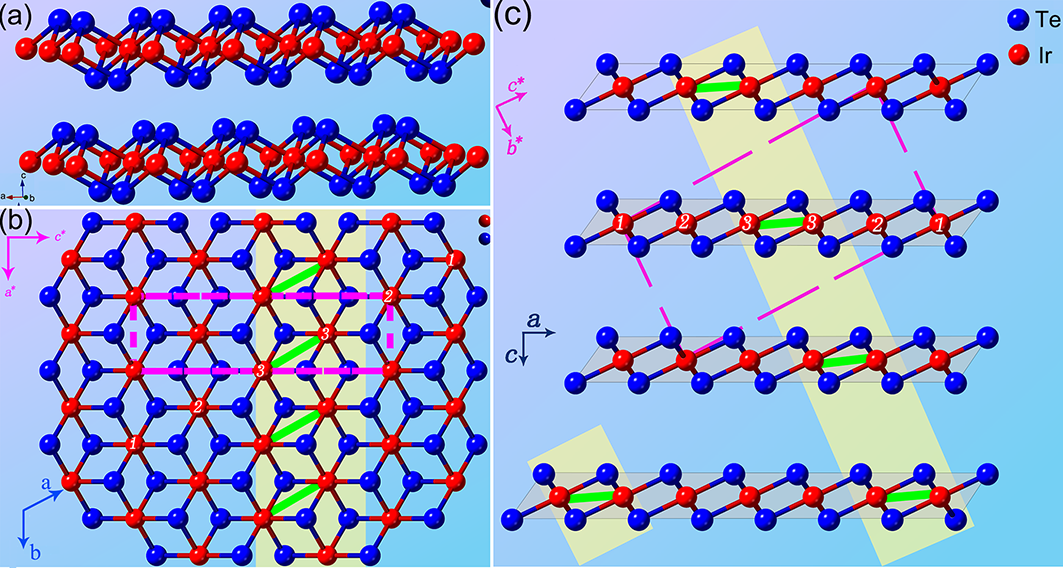}
\caption{(a)The crystal structure of IrTe$_2$ at the room temperature (RT). The crystal structure of IrTe$_2$ below the 1/5 dimerization: (b) the triangular Ir and Te layers and (c) the layers stacking along c axis.  (a, b, and c) and  (a$^\ast$, b$^\ast$, and c$^\ast$) are the crystallographic axes at high and low temperature,respectively. The pink box represents the unit cell of the 1/5 phase, and the numeric labels denote the inequivalent Ir sites. The Ir(3)-Ir(3) dimers and their stripes are marked with green bonds and yellow shading, respectively.}
\end{figure*}

\begin{figure*}[htbp]
\centering
\includegraphics[width=17.2cm]{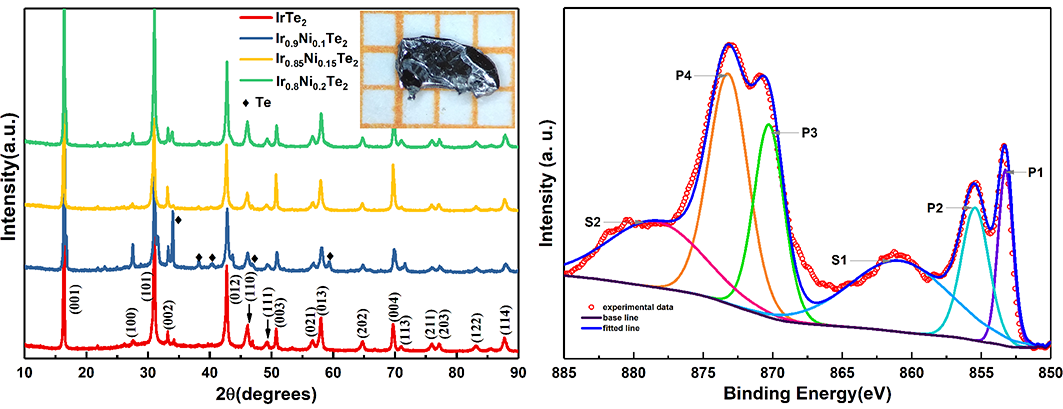}
\caption{(a)The powder XRD pattern of Ir$_{1-x}$Ni$_x$Te$_2$ at room temperature. The inserted part is the crystal morphology of Ir$_{0.8}$Ni$_{0.2}$Te$_2$. The background scale is in millimeters for the single crystal. (b) The normalized core-level photoemission spectrum of Ni in Ir$_{0.8}$Ni$_{0.2}$Te$_2$. }
\end{figure*}
\begin{figure*}[htbp]
\centering
\includegraphics[width=17.2cm]{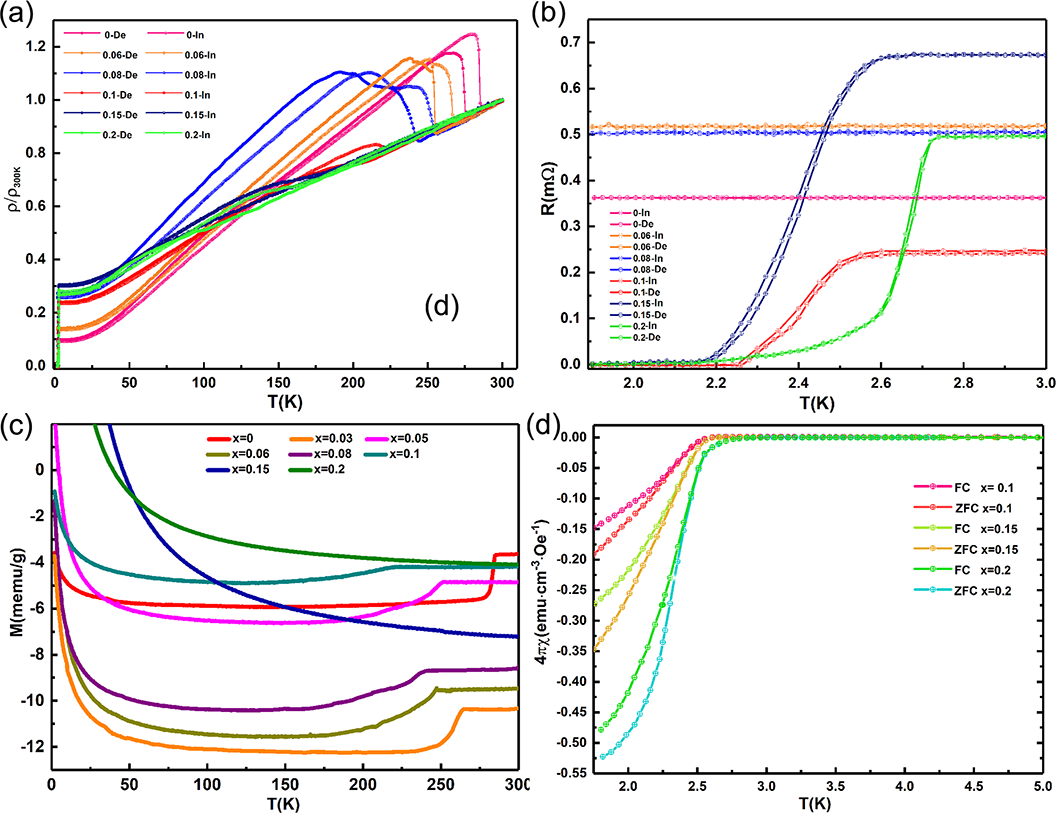}
\caption{(a)The temperature dependence of normalized resistivity$\rho$/$\rho$$_{300K}$ for Ir$_{1-x}$Ni$_x$Te$_2$ (x=0, 0.06, 0.08,0.1, 0.15 and 0.2). Here, the curve with “De” was obtained during the decreasing temperature process, while the curve with “In” was obtained during the increasing temperature process. (b) shows the superconducting transition part of Ir$_{1-x}$Ni$_x$Te$_2$. The double lines correspond to the lines gotten during the increasing and decreasing temperature process, respectively. (c) The temperature dependent DC magnetic susceptibility of Ir$_{1-x}$Ni$_x$Te$_2$ with x=0, 0.03, 0.05, 0.06, 0.08, 0.1, 0.15 and 0.2, respectively, from 2 K to 300 K under a field of 2 T, parallel to ab-plane. (d) Low temperature DC magnetic susceptibility 4$\pi$$\chi$ versus temperature for Ir$_{1-x}$Ni$_x$Te$_2$ with x=0.1,0.15 and 0.2, respectively, obtained in a magnetic field of 30 Oe, parallel to ab-plane.}
\end{figure*}
\begin{figure*}[t]
\centering
\includegraphics[width=17.2cm]{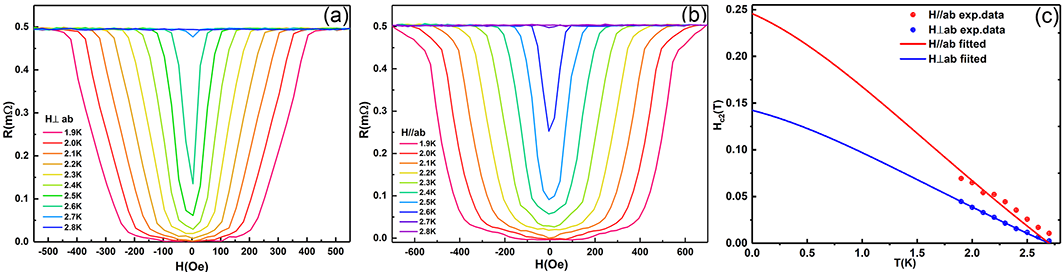}
\caption{Determination of the upper critical magnetic fields for Ir$_{0.8}$Ni$_{0.2}$Te$_2$. (a) and (b) Resistance as a function of magnetic field applied out of  and in ab plane from 1.9 to 2.8 K. (c)Extracted upper critical magnetic fields H$_{c2}$ for H//ab and H$\perp$ab as a function of temperature. The solid line is the fitted line according to WHH theory.}
\end{figure*}
\begin{figure}[htbp]
\centering
\includegraphics[width=8.6cm]{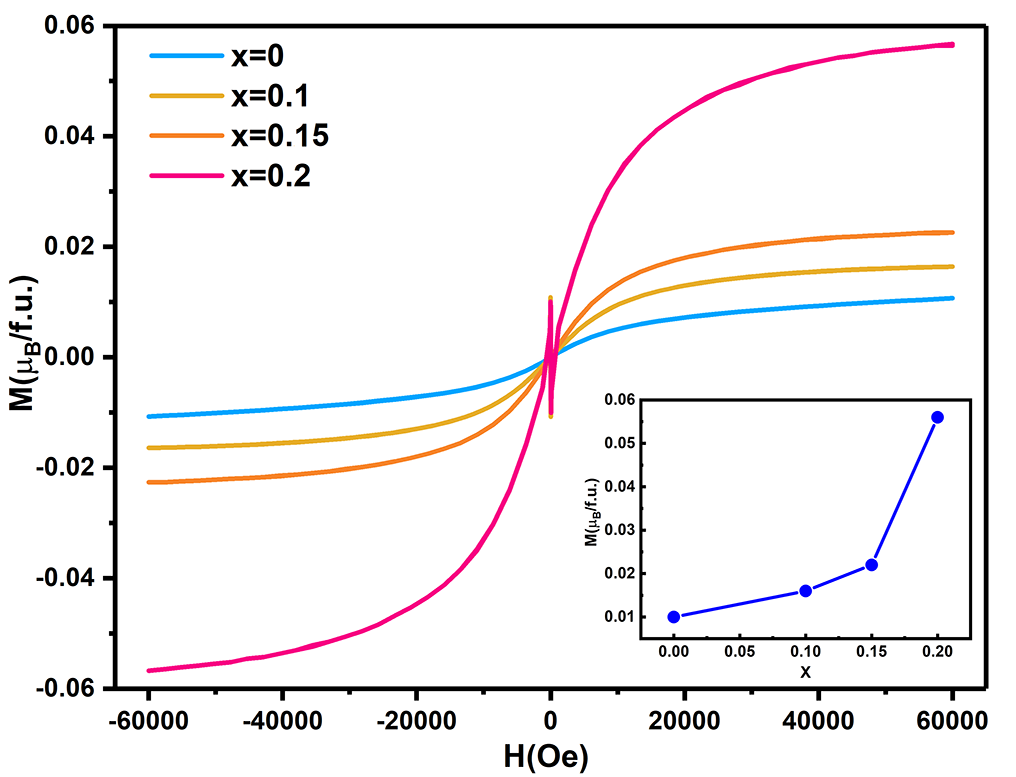}
\caption{M-H loop of Ir$_{1-x}$Ni$_x$Te$_2$ at 1.8K under applied magnetic fields up to 6T.}
\end{figure}
\begin{figure*}[htbp]
\centering
\includegraphics[width=17.2cm]{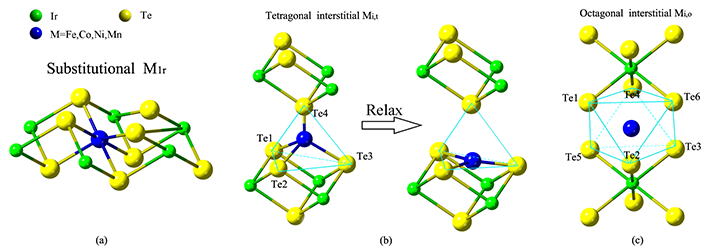}
\caption{Local atomic geometries of (a) substitutional M$_{Ir}$,  (b) unrelaxed M$_{i,t}$, (c) relaxed M$_{i,t}$, and (d) M$_{i,o}$.}
\end{figure*}
\begin{figure}[htbp]
\centering
\includegraphics[width=8.6cm]{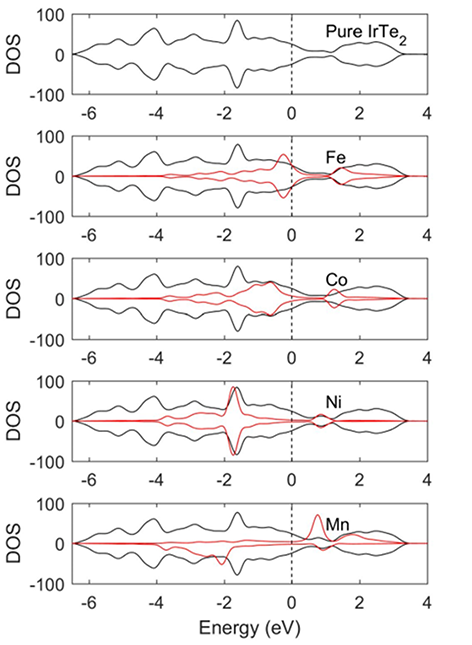}
\caption{Calculated density of states (DOSs). Black lines denote total DOS, red ones denote the projected DOSs of the substitutional magnetic atoms.
}
\end{figure}
\begin{figure}[t]
\centering
\includegraphics[width=8.6cm]{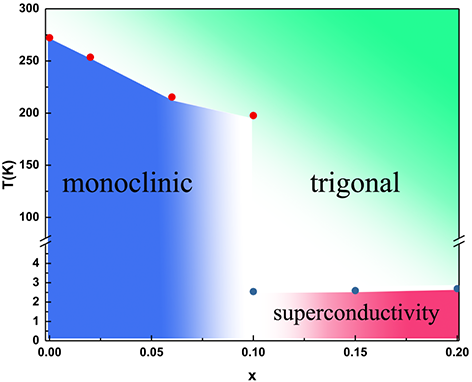}
\caption{Electronic phase diagram of Ir$_{1-x}$Ni$_x$Te$_2$ as a function of x. Red symbols correspond to the first order structural phase transition temperatures collected during heating process, and blue symbols represent the superconducting transition temperatures. The superconductivity and structural phase transition coexist at x=0.1}
\end{figure}
Single crystals of Ir$_{1-x}$Ni$_x$Te$_2$ have been successfully grown via self-flux method. Ir, Ni and Te powder was mixed in an atomic ratio of (1-x): x: 5, then was placed in an Al$_2$O$_3$ crucible and sealed in an evacuated quartz tube. The mixture was initially heated up to 950$^{\circ}$C and kept for several hours, then to 1160$^{\circ}$C for one day, and finally cooled down slowly to 900$^{\circ}$ C at a rate of 2$^{\circ}$C/h. At last, the remanent Te was separated from single crystals by centrifuging. The DC transport data was collected by the  PPMS of Quantum Design company with a four-probe method. The magnetic susceptibility was performed in a quantum design superconducting quantum interference device vibrating sample magnetometer system (SQUID-VSM). XPS was performed on Axis Ultra of Kratos Analytical with AlK$\alpha$ (h$\nu$=1486.6eV) as the monochromatic X-ray radiation source. The power of the X-ray was about 150 W. All peaks had been fitted using a Shirley background and Voigt (mixed Lorentzian-Gaussian) line shapes. First-principles calculations were performed with the PBE pseudopotentials \cite{65} as implemented in the VASP codes \cite{66}. The van de Waals (vdW) correction was involved using the optB88-vdW scheme\cite{67}, and spin polarization was taken into account. More details have been listed in the supplementary materials.
\section{Result and Discussion}

As shown by Fig. 1(a) that, at room temperature IrTe$_2$ has a layered structure with the space group of P$\overline{3}$m1 belonging to trigonal system. Ir ions are coordinated octahedrally with six Te ions. Differently from typical layered transitional metal dichalcogenides, IrTe$_2$ layers bond with each other through Te-Te bonding rather than weak van der Waals force\cite{50,51}.  Below the phase transition temperature, IrTe$_2$ would enter into monoclinic system \cite{52}. As illustrated in Fig. 1(b), the 1/5 phase is stable at lower temperature \cite{42,43,44,46}. In this phase, Ir-Ir dimerizes between Ir(3)-Ir(3) sites along the a axis together with Te-Te dimerization (not shown), accompanied by charge transfer from Ir to Te atoms \cite{43,45,46}. The Ir ions with a relatively larger valence (Ir$^{4+}$ like) form dimers running along the b axis (a$^\ast$ axis), while those with smaller valence (Ir$^{3+}$like) do not participate in dimerization. As shown by Fig. 1 (c) that, the dimerization leads to stripes of Ir atoms with different charges, and the sequence along the a axis is  “33344,” where “3”(“4”) denotes the Ir$^{3+}$like (Ir$^{4+}$like) site \cite{46}. In the neighboring layer, the location of these stripes shifts by a unit cell, resulting in a staircaselike arrangement with a modulation vector q$_1$=(1/5, 0, -1/5)\cite{46}.\par
\begin{table}
\caption{\label{tab:table2}The lattice parameters of Ir$_{1-x}$Ni$_x$Te$_2$ }
\begin{ruledtabular}
\begin{tabular}{ccc}
 &a(\r{A})&c(\r{A})\\
\hline
IrTe$_2$&3.93055 & 5.38946  \\
Ir$_{0.9}$Ni$_{0.1}$Te$_2$& 3.92853 & 5.38401   \\
Ir$_{0.85}$Ni$_{0.15}$Te$_2$& 3.9269& 5.38399  \\
Ir$_{0.8}$Ni$_{0.2}$Te$_2$& 3.92232 & 5.36866 \\
NiTe$_{2}$& 3.8550 & 5.2659 \cite{69} \\
\end{tabular}
\end{ruledtabular}
\end{table}

Fig.2 (a) shows the powder X-ray diffraction patterns of Ir$_{1-x}$Ni$_x$Te$_2$, with nominal composition, at room temperature. The Ir$_{1-x}$Ni$_x$Te$_2$ powder were gotten by smashing the single crystals. The main observed reflections could be indexed according to the P$\overline{3}$m1 space group of trigonal system. The other impurity peaks mainly come from flux agent tellurium. Table\newcommand{\RNum}[1]{\uppercase\expandafter{\romannumeral #1\relax}} \RNum{1} lists the refined lattice parameters of Ir$_{1-x}$Te$_x$Ni$_2$. It shows that both a axis and c axis shrink slightly with the increase of the doping amount of Ni, which coincides with the fact that both the lattice parameters of a axis and c axis of NiTe$_2$ are smaller than those of IrTe$_2$. Fig. 2(b) shows two spin-orbit doublets and two shakeup satellites (labelled as “S1”\&“S2”) of Ni element in Ir$_{0.8}$Ni$_{0.2}$Te$_2$. Peak 1 and Peak 3 centered at 853.3eV and 870.3 eV, respectively, coinciding with the binding energy of Ni$^{2+}$ in NiTe$_2$\cite{47,48}, which means that the Ni atoms have substituted for Ir atoms successfully. Peak 2 and Peak 4 centered at 853.3eV and 870.3 eV, respectively, can be designated to Ni$^{2+}$ arising from inevitable surface oxidation of Ir$_{0.8}$Ni$_{0.2}$Te$_2$\cite{47,49}. S1 and S2 locating at the position of 861.00eV and 878.80eV are their corresponding shakeup satellites, respectively. \par
The structural transition temperature T$_s$ could be determined from both transport and magnetic measurements. Fig. 3 (a) shows the normalized dc resistivity $\rho$/$\rho$$_{300K}$ of Ir$_{1-x}$Ni$_x$Te$_2$ over the temperature range of 2 K-300 K. For pure IrTe$_2$, steep jumps at 275 K (reaching maximum at 262 K) on cooling and 286 K (maximum at 278 K) on heating are observed, respectively. The significant hysteresis could be ascribed to a first order phase transition. The resistivity $\rho$(T) keeps decreasing with lowering temperature after the phase transition. However, similar with the transport behaviour of Ir$_{1-x}$Pt$_x$Te$_2$, an upward curvature is observed below 50 K, which may be due to electron-phonon interactions \cite{18}. The doping of Ni could suppress the structural phase transition. It could be observed from Fig.3 (a) and (c) that, the transition temperature T$_s$ decreases with the increase of the doping amount. The structural phase transition is almost suppressed at x=0.2 according to the transport result. With the suppression of the structural phase transition, the superconducting phase emerges for samples with x$\geq$ 0.10. The superconductivity and structural phase transition coexist
at x=0.1. Even no superconducting transition is observed for pure IrTe$_2$ in the transport measurement, when a magnetic field parallel to ab-plane is applied, a weak diamagnetic signal is observed (see Fig.S1(a) in the Supplemental Materials). It could be ascribed to the filament superconductivity, possibly originated from Ir vacancies or excess Te in the samples \cite{18,61}. The diamagnetic signal could also be observed in Ir$_{0.94}$Ni$_{0.06}$Te$_2$ and Ir$_{0.92}$Ni$_{0.08}$Te$_2$ samples (see Fig.S1(b) and (c)), but they are not the bulk superconductivity according to the transport measurement and superconductivity volume calculation result. Bulk superconductivity could be observed in Ir$_{1-x}$Ni$_x$Te$_2$ samples with x=0.1, 0.15 and 0.2, as shown by Fig.3(b) and (d). Their onset superconducting transition temperatures are 2.55K, 2.6K and 2.7K, respectively. As shown by Fig.3(d) that, the onset diamagnetic transition temperatures of three samples coincide well with the transport result. The upper limit of the superconducting volume fraction, measured in the ZFC process, reaches 19$\%$, 35$\%$ and 52$\%$, respectively, which means that they are bulk superconductivity. Simultaneously, the minor distinction between FC and ZFC processes implies weak superconducting vortex pinning, which indicates the good quality of our single crystals. Fig.3 (c) shows the magnetic susceptibilities of Ir$_{1-x}$Ni$_x$Te$_2$ single crystals in an applied magnetic field H=1T parallel to ab–plane. For samples with Ni-doping amount of 0 $\leq$ x $\leq$ 0.1, the diamagnetism which is in independent of the temperature is observed above their structural phase transitions, and then drops suddenly at the transition temperatures, resulting in much more diamagnetism. For Ir$_{0.85}$Ni$_{0.15}$Te$_2$ and Ir$_{0.8}$Ni$_{0.2}$Te$_2$ samples, the sudden drop of the magnetic susceptibility marking the phase transition has disappeared, which coincides with the transport result, as the resistivity jump becomes such even too. However, compared with the magnetic result, the resistivity jump marking the phase transition vanishes hysteretically. It could be due to the fact that the transport measurement is more sensitive to the structural phase transition or the field applied to the samples is not large enough. The upturn of  the magnetic susceptibility at low temperature is likely from paramagnetic impurities. \par
Fig.4(a) and (b) show the resistance of Ir$_{0.8}$Ni$_{0.2}$Te$_2$ as a function of magnetic field applied out of (H$\perp$ab) and in ab plane (H//ab) from 1.9 to 2.8 K, respectively. It could be observed from the curves that the superconductivity of Ir$_{0.8}$Ni$_{0.2}$Te$_2$ shows the anisotropic behaviour. The superconducting state is much more robust against in-plane magnetic field (H//ab) than the out-of-plane magnetic field (H$\perp$ab).  In Fig.4(a) and (b), the upper critical field H$_{c2}$ is defined as the superconducting onset temperature. Fig. 4(c) summarizes the upper critical field H$_{c2}$ as a function of temperature. The linear temperature dependence close to T$_c$ is obtained for H$_{c2}$, suggesting the dominance of only one type of bulk carrier. The solid line is the fitted line based on Werthamer-Helfand-Hohenberg (WHH) theory. We obtained the upper critical field at zero temperature 	H$_{c2,\perp}$(0)=0.14 T and H${_{c2, //}}$(0) = 0.24 T.Based on H$_{c2,\perp}$, the coherence length $\xi$$_{//}$=$\sqrt{{\Phi}_{0}/{2{\pi}H_{{c2},\perp}}}$= 48 nm is obtained, while $\xi$$_{\perp}$=28 nm is gotten according to
$\xi$$_{//}$$\xi$$_{\perp}$=${\Phi}_{0}/{2{\pi}H_{{c2,//}}}$.Then the dimensionless anisotropy parameter $\gamma$=$\xi$$_{//}$/$\xi$$_{\perp}$$\sim$ 2,which is similar with the reported value of Ir$_{0.95}$Pd$_{0.05}$Te$_2$\cite{50}.\par
In order to confirm its magnetic property, the M-H loop of Ir$_{1-x}$Ni$_x$Te$_2$ is measured at 1.8K. Neither ferromagnetic hysteresis loop nor obvious remanence is observed from Fig.5, which means that Ir$_{1-x}$Ni$_x$Te$_2$ does not show ferromagnetic order at 1.8K. The diamagnetic behaviour of Ir$_{1-x}$Ni$_x$Te$_2$ (x$\neq$0) under low field could be ascribed to the superconductive diamagnetism. All of them demonstrate paramagnetic behaviour and show saturated at high field, with the saturation magnetization of 0.010, 0.016, 0.022 and 0.056 $\mu$$_B$/f.u. for x=0, 0.1, 0.15 and 0.2, respectively.  As shown by the inset of Fig.5(a) that below x$\leq$ 0.15, the saturation magnetization versus x shows almost linear behaviour, while when x=0.2, it deviates largely. This could be interpreted as following: the paramagnetism may be from the tiny amount of Ni atoms that have not taken up the position of Ir as substitution doping (the substitution doping of Ni brings no magnetism which is elaborated in the calculation part ). Furthermore, according to the inductively coupled plasma (ICP) result,  the maximal doping ratio of Ni in IrTe$_2$ single crystals is around 20\% i.e. x=0.2. With x=0.2, the Ni atoms that could not occupy the substitution position increase dramatically which makes its magnetization deviate the linear behaviour largely.\par
To investigate the most possible doping position of M (M=Mn, Fe, Co and Ni), the substitutional impurities on the Ir site, interstitials on the tetragonal and octagonal sites (see Fig. 6) are considered. The calculated formation energies in Ir-rich and Te-rich conditions are listed in Table S1 and Table S2 of the supplementary materials, respectively, and the magnetic moments of the doped supercells are listed in Table S3. According to calculations, the tetragonal interstitials are always not stable. Impurity atoms initially placed on the tetragonal site spontaneously relax downward to the Te triangle center, as shown by Fig.6 (b). On the contrary, they could stay on the octagonal site after relaxation. All interstitials have high formation energy, ranging from 0.15 to 1.31 eV, thus they are very unlikely to occur in considerable equilibrium concentrations. Co, Ni and Mn impurities energetically prefer to substitute on the Ir site from Te-rich to Ir-rich growth conditions. Fe impurities prefer the Ir site in Te-rich conditions, but they are energetically more favorable on the octagonal interstitial site in Ir-rich conditions. The substitutional Fe$_{Ir}$ and Co$_{Ir}$ always have high formation energy, hence high equilibrium concentrations cannot be expected. The substitutional Ni$_{Ir}$ and Mn$_{Ir}$ have very low formation energy. This is true especially in Te-rich conditions. Both impurities have negative formation energy, meaning that they can be easily formed in equilibrium conditions. From Te-rich to Ir-rich conditions, the formation energy increases. But even in extremely Ir-rich conditions, the formation energy of Ni$_{Ir}$ is just slightly above zero (0.02 eV). Obviously, Te-rich conditions are energetically more favorable for the substitutional doping of Ni and Mn. To study the electronic structure of Ir$_{1-x}$Ni$_x$Te$_2$ (M = Mn, Fe, Co, and Ni), the density of states is plotted in Fig. 7. Here, the Fermi level is taken as reference. For Fe, Co and Ni doped IrTe$_2$, their up and down spin density of states compensate each other well, which makes them show no magnetism. However, for Mn doped IrTe$_2$,the up and down spin density of states do not compensate each other which introduces a magnetic moment of 3.0$\mu$$_B$ to the supercell.\par
Our results are summarized in the electronic phase diagram presented in Fig. 8. The structural phase transition temperature of IrTe$_2$ is strongly suppressed with Ir substitution. As soon as the phase transition is significantly suppressed, the superconductivity appears. The superconducting state appears at x=0.1, and superconducting T$_c$ reaches the maximum of 2.7 K for x=0.2. The structural phase transition and superconductivity coexist at x=0.1.

\section{Conclusions}
In conclusion, in this study, Ir$_{1-x}$Ni$_x$Te$_2$ single crystals have been synthesized. The Ni atoms have substituted for Ir in  Ir$_{1-x}$Ni$_x$Te$_2$. Like the CDW behaviour, the structural phase transition shows competition and coexistence with the superconductivity. The monoclinic phase transition has been suppressed gradually with the increase of Ni-doping amount, giving rise to the stabilization of the trigonal phase with superconductivity. Within 0.1$\leq$x$\leq$0.2, Ir$_{1-x}$Ni$_x$Te$_2$ shows the superconductive behaviour with T$_c$ around 2.6K. The superconductivity shows anisotropy. Ir$_{1-x}$Ni$_x$Te$_2$ only shows weak paramagnetism and no ferromagnetic order is observed in it, which could be ascribed to the well compensation of their up and down spin density of states. In addition, according to the calculation result, Co, Ni and Mn impurities energetically prefer to substitute on the Ir site from Te-rich to Ir-rich growth conditions. Fe impurities prefer the Ir site in Te-rich conditions, but they are energetically more favorable on the octagonal interstitial site in Ir-rich conditions.  For Fe, Co and Ni doped IrTe$_2$, they do not show magnetism. However, for Mn doped IrTe$_2$, it has a magnetic moment of 3.0$\mu$$_B$ to the supercell.

\begin{acknowledgments}
This work was supported by the Ministry of Science and Technology of China (Grant Nos. 2016YFA0301003 and 2016YFA0300501) and the National Natural Science Foundation of China (NNSFC) (Grant Nos. U1632272, 11574201, 11804194,U1632102, 11674224, 11974243, 11774223, U1732154, and 11521404). D.Q. acknowledges support from the Changjiang Scholars Program. D.Q., J.M., and W.Z. acknowledge additional support from a Shanghai talent program.
\end{acknowledgments}

\bibliographystyle{unsrt}
\bibliography{MS-3}

\end{document}